\documentclass[twocolumn,showpacs,preprintnumbers,amsmath,amssymb]{revtex4}
\usepackage{graphicx}% Include figure files
\usepackage{dcolumn}% Align table columns on decimal point
\usepackage{bm}% bold math

\begin{document}

%\preprint{APS/123-QED}

\def\hm{\ \rm {\it h}^{-1} Mpc}

\def\ltsim{\, \lower2truept\hbox{${<
      \atop\hbox{\raise4truept\hbox{$\sim$}}}$}\,}

\def\gtsim{\, \lower2truept\hbox{${>
      \atop\hbox{\raise4truept\hbox{$\sim$}}}$}\,}

\def\omde{\Omega_{\rm DE}}
\def\cs2{c_s^2}

\title{Integrated Sachs-Wolfe effect from the cross correlation of WMAP3 year and the NRAO VLA sky survey data: New results and constraints on dark energy}% Force line breaks with \\

\author{Davide Pietrobon$^1$}\email{davide.pietrobon@roma2.infn.it}
\author{Amedeo Balbi$^2$}\email{amedeo.balbi@roma2.infn.it}
\author{Domenico Marinucci$^3$}\email{marinucc@mat.uniroma2.it}
\affiliation{$^1$Dipartimento di Fisica, Universit\`a di Roma ``Tor Vergata'',  V. della Ricerca Scientifica 1, I-00133 Roma, Italy\\
$^2$Dipartimento di Fisica, Universit\`a di Roma ``Tor Vergata''
and INFN Sezione di Roma ``Tor Vergata'', V. della Ricerca Scientifica 1, I-00133 Roma, Italy\\
$^3$Dipartimento di Matematica, Universit\`a di Roma ``Tor Vergata'', V. della Ricerca Scientifica 1, I-00133 Roma, Italy }

\date{\today}% It is always \today, today,
             %  but any date may be explicitly specified

\begin{abstract}
We cross-correlate the new 3 year Wilkinson Microwave Anistropy Probe (WMAP)  cosmic microwave background (CMB) data with the NRAO VLA Sky Survey (NVSS) radio galaxy data, and find further evidence of late integrated Sachs-Wolfe (ISW) effect taking place at late times in cosmic history. Our detection makes use of a novel statistical method \cite{Baldi et al. 2006a, Baldi et al. 2006b} based on a new construction of spherical wavelets, called needlets. The null hypothesis (no ISW) is excluded at more than 99.7\% confidence. When we compare the measured cross-correlation with the theoretical predictions of standard, flat cosmological models with a generalized dark energy component parameterized by its density, $\omde$, equation of state $w$ and speed of sound $\cs2$, we find $0.3\leq\omde\leq0.8$ at 95\% c.l., independently of $\cs2$ and $w$. If dark energy is assumed to be a cosmological constant ($w=-1$), the bound on density shrinks to $0.41\leq\omde\leq 0.79$. Models without dark energy are excluded at more than $4\sigma$. The bounds on $w$ depend rather strongly on the assumed value of $\cs2$. We find that models with more negative equation of state (such as phantom models) are a worse fit to the data in the case $\cs2=1$ than in the case $\cs2=0$.
\end{abstract}

% PACS, the Physics and Astronomy
                             % Classification Scheme.
%\keywords{Suggested keywords}%Use showkeys class option if keyword
                              %display desired
\maketitle

\section{Introduction}

The most outstanding problem in modern cosmology is understanding the mechanism that led to a recent epoch of accelerated expansion of the universe. The evidence that we live in an accelerating universe is now compelling. The luminosity distance at high redshift ($z\sim 1$) measured from distant type Ia supernovae is consistent with a negative deceleration parameter ($q_0<0$ at $\sim 3\sigma$) and shows strong evidence of a recent transition from deceleration to acceleration \cite{Riess et al. 2004, Riess et al. 1998, Perlmutter et al. 1999}. The amount of clustered matter in the universe, as detected from its gravitational signature through a variety of large scale probes (redshift surveys, clusters of galaxies, etc.)  cannot be more than $\sim 1/3$ of the total content of the universe \cite{Springel et al. 2006}. Observations of the cosmic microwave background (CMB) anisotropy have constrained the value of cosmological parameters with outstanding precision. The recent WMAP data (\cite{Bennett et al. 2003, Spergel et al. 2003, Spergel et al. 2006}) have shown that the total density of the universe is very close to its critical value. Taken together, these results are a strong indication in favor of a non-null cosmological term, which would at the same time explain the accelerated expansion of the universe and provide the remaining $\sim 2/3$ of its critical density.

The precise nature of the cosmological term which drives the accelerated expansion, however, remains mysterious. The favoured working hypothesis is to consider a dynamical, almost homogeneous component (termed {\em dark energy}) with negative pressure (or, equivalently, repulsive gravity) and an equation of state $w\equiv p/\rho<-1/3$ \cite{Peebles & Ratra 1988, Caldwell et al. 1998, Wang et al. 2000, Peebles & Ratra 2003}. Such a framework helps alleviating a number of fundamental problems arising when a constant cosmological term is interpreted as the energy density of the vacuum \cite{Weinberg 1989}.

One key indication of an accelerated phase in cosmic history is the signature from the integrated Sachs-Wolfe (ISW) effect \cite{Sachs & Wolfe 1967} in the CMB angular power spectrum. This is directly related to variations in the gravitational potential: in particular, it traces the epoch of transition from a matter-dominated universe to one dominated by dark energy. This effect (which is usually called {\em late ISW}, as opposed to a {\em early ISW} generated during the radiation-matter transition), shows up as a contribution in the low multipole region of the CMB spectrum. A detection of a late ISW signal in a flat universe is, in itself, a direct evidence of dark energy. Furthermore, the details of the ISW contribution depend on the physics of dark energy, and are therefore a powerful tool to better understand its nature. Unfortunately, the low multipole region of the angular power spectrum is also  the most affected by cosmic variance, making the extraction of the ISW signal a difficult task.

A useful way to separate the ISW contribution from the total signal is to cross-correlate the CMB anisotropy pattern (imprinted during the  recombination epoch at $z\sim 1100$)  with tracers of the large scale structure (LSS) in the local universe \cite{Crittenden & Turok 1996}. Detailed predictions of the ability to reconstruct the ISW using this technique were obtained by a number of authors \cite{Cooray  2002, Hu & Scranton 2004, Afshordi 2004, Corasaniti et al. 2003, Pogosian et al. 2005}. This kind of analysis has been performed several times during the past few years, using different CMB data sets and various tracers of clustering. The first detection of the ISW \cite{Boughn & Crittenden 2004, Boughn & Crittenden 2005} was obtained  by combining the WMAP 1st year CMB data with the hard X-ray background observed by the High Energy Astronomy Observatory-1 satellite (HEAO-1 \cite{Boldt 1987}) and with the radio galaxies of the NRAO VLA Sky Survey (NVSS \cite{Condon et al. 1998}). The positive correlation with NVSS was later confirmed by the WMAP team \cite{Nolta et al. 2004}. Other large scale structure tracers that led to similar positive results were the APM galaxy survey \cite{Maddox et al. 1990}, the Sloan Digital Sky Survey (SDSS \cite{York et al. 2000}) and the near infrared 2 Micron All Sky Survey eXtendend Source Catalog (2MASS XSC \cite{Jarrett et al. 2000}) \cite{Fosalba et al. 2003, Scranton et al. 2003, Fosalba & Gaztanaga 2004, Afshordi et al. 2004, Padmanabhan et al. 2005, Cabre et al. 2006}.

A somewhat different strategy to attack the problem was recently adopted by other authors, who attempted to seek the ISW signal in spaces other than the pixel space of the maps or the harmonic space of the angular power spectrum \cite{Vielva et al. 2006, McEwen et al. 2006}. This approach relies on spherical wavelets as a tool to exploit the spatial localization of ISW (at large angular scales) in order to get a more significant detection of the effect. 

The purpose of the present paper is twofold. On one side, we want to perform a further analysis of the CMB-LSS cross-correlation, in order to obtain an independent check on previous results. We combine the recent 3rd year release of WMAP CMB sky maps with the radio galaxy NVSS catalogue, and carry out our investigation in wavelet space. We make use of a new type of spherical wavelets, the so-called {\em needlets} \cite{Narcowich et al. 2006, Baldi et al. 2006a, Baldi et al. 2006b} which have a series of advantages over previously used wavelets, as will be described in detail later. This then represents at the same time a check on previous results \cite{Vielva et al. 2006, McEwen et al. 2006} and a significant improvement of the statistical and technical aspects of the problem. On the other side, we follow a rather general approach to dark energy modelization, as first proposed in \cite{Hu 1998}. Within this framework the phenomenology of dark energy is characterized by three physical parameters: its overall density $\omde$,  its equation of state $w$, and the sound speed $c_s^2$. This parameterization has the advantage of being model independent, allowing one to encompass a rather broad set of fundamental models, and of giving a more realistic description of the dark energy fluid, for example accounting for its clustering properties, a feature that was shown to have quite a strong effect on theoretical predictions \cite{Weller & Lewis 2003}. As shown in \cite{Hu & Scranton 2004, Bean & Dore 2004, Corasaniti et al. 2003} the ISW signature can in principle be able to set constraints on the parameters of this generalized dark energy scenario: however, previous analyses of the ISW from CMB-LSS cross-correlation made a number of unrealistic simplyfing assumptions on the dark energy component and were only able to either found confirmations for its existence by constraining its density, or to set limits on its equation of state under restrictive hypotheses on its clustering properties (one notable exception being the analysis performed in \cite{Corasaniti et al. 2005} which applied a parameterization similar to ours to make a likelihood analysis of the cross-correlation data points estimated in \cite{Gaztanaga et al. 2004}). Our approach is more ambitious, as we attempt a more realistic description of dark energy and derive constraints on the combined set of three above mentioned parameters. 

The paper is organized as follows. In Section \ref{data} we describe the dataset we use to perform our analysis. In Section \ref{techniques} we  outline the theoretical predictions of the expected ISW signal and discuss the statistical techniques we apply to extract it from the data. In Section \ref{results} we present our results and the derived constraints on dark energy. Finally, in Section \ref{conclusions} we discuss our main findings and conclusions.

\section{Data}
\label{data}

We trace the local distribution of matter in the universe by using the NVSS radio galaxy catalogue \cite{Condon et al. 1998}. This dataset contains roughly $1.8\times 10^6$ point sources observed at 1.4 GHz. The flux limit of the catalogue is at $\sim 2.5$ mJy, resulting in a completeness of about $50\%$. The survey covers about $80\%$ of the sky, at $\delta>-40^\circ$. We construct a point source map (after removal of about $3\times 10^5$ resolved sources) using the $N_{\rm side}=64$ HEALPix pixelization \cite{Gorski et al. 2005}. Such a map has $49\,152$ pixels of about 1 degree side, and guarantees a good sampling of source counts in each pixel. We conservatively exclude from the map all sources with $\delta>-37^\circ$ since the coverage becomes very poor when approaching that value of declination. The final map we use has roughly 35 sources per pixels on average. It was pointed out \cite{Boughn & Crittenden 2002} that there is a declination dependence of the mean source density in the catalogue, since the survey had different integration time in some well-defined constant-declination bands on the sky. As suggested in previous analyses \cite{Nolta et al. 2004, Vielva et al. 2006} we correct for this spurious effect by subtracting the average source count in each constant-declination band. 

Our CMB dataset consists of the internal linear combination (ILC) temperature map from the 3 year release of WMAP\cite{lambda}. This map is produced by combining 5 smoothed temperature maps, with weights chosen in such a way to produce minimal Galactic foreground contamination while mantaining the CMB signal. According to the WMAP team \cite{Hinshaw et al. 2006, Nolta et al. 2004} the ILC map gives a reliable estimate of the CMB signal at low multipoles with negligible instrumental noise. We believe this is appropriate with respect to our goals, since the late ISW effect is expected to peak exactly at such large angular scales. As an additional caution, we mask out the Galactic plane region of the map and bright point sources using one of the templates produced by the WMAP team, namely the conservative Kp0 intensity mask \cite{lambda, Hinshaw et al. 2006}. 

While the original map ILC was produced at a resolution $N_{\rm side}=512$ in the HEALPix pixelization scheme \cite{Gorski et al. 2005} (consisting of 12$N_{\rm side}^2=3\,145\,728$ pixels) we degrade it to a resolution of $N_{\rm side}=64$ to match the resolution of the NVSS. This resolution is appropriate for the CMB as well, since we are not interested in the fine-scale details of the map. 

A joint mask, including both the Kp0 mask and the NVSS declination limit, is applied to both maps used in the analysis.

\begin{figure}
\includegraphics[width=\columnwidth]{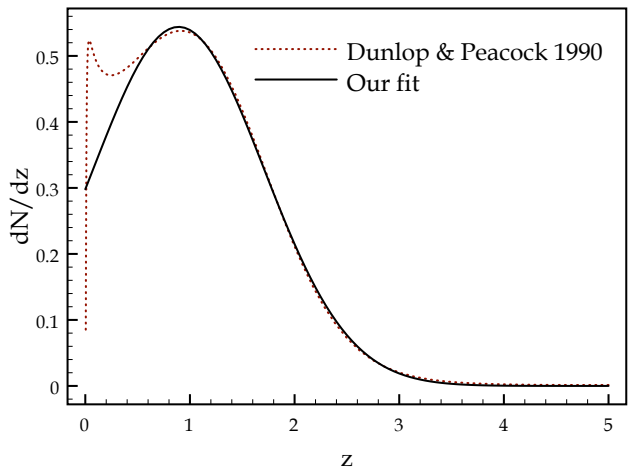}
\caption{The function dN/dz used for the sources in the NVSS in our analysis. The dotted curve is the theoretical model from \cite{Dunlop & Peacock 1990}, which has a spurious feature due to the breakdown of the fit at low z. The continuous line is the fit adopted in our analysis.\label{fig:dndz}}
\end{figure}

There is no redshift information for the individual sources in the catalogue. Nonetheless, some knowledge of the $dN/dz$ function is needed to connect the observed source count fluctuation $\delta n$ to the underlying matter fluctuation $\delta \rho$ (as we will show later). We then use a fit to the $dN/dz$ estimated in \cite{Dunlop & Peacock 1990} and already applied to previous analysis of the NVSS catalog \cite{Nolta et al. 2004}. Since the fit in \cite{Dunlop & Peacock 1990} breaks down at low redshifts, we have approximated it with a Gaussian $dN/dZ$ centered around $z\simeq 0.9$ with a width $\Delta z \simeq 0.8$, and normalized in order to give a unit integral. The resulting $dN/dz$ used in our analysis is shown in Figure \ref{fig:dndz}. We have verified that the difference at low z has negligible effect on the final results.

\section{Techniques}
\label{techniques}

\subsection{Cross-correlation between CMB and LSS}

It is common practice to expand the map of the CMB temperature fluctuations into spherical harmonics ($Y_{lm}$) as:
\begin{equation}
\delta T=\sum_{lm} a_{lm}^T Y_{lm}(\theta,\phi)
\end{equation}
in order to extract the angular power spectrum:
\begin{equation}
C^{TT}_l=\langle\vert a^{T}_{lm}\vert^2\rangle
\end{equation}
which enters in the two-point auto-correlation function of the CMB as:
\begin{eqnarray}
C^{TT}(\alpha)&\equiv&\langle \delta T_1\delta T_2\rangle=\nonumber\\ 
&=&\sum_l {(2l+1)\over 4\pi}B^2_{T,l}C^{TT}_l P_l(\cos\alpha)
\end{eqnarray}
where $P_l$ are the Legendre polinomials, $\alpha$ is the angular separation between two given points, and the function $B_{T,l}$ models the experimental beam response and the pixel window function of the map.

In an equivalent way, given a projected source count map: 
\begin{equation}
\delta n=\int dz\, b(z){dN\over dz}\delta(z)
\end{equation}
(where $\delta$ is the underlying matter fluctuation in a given direction, $b$ is the bias parameter, and $dN/dz$ was discussed previously) we can define the source count auto-correlation function:
\begin{eqnarray}
C^{NN}(\alpha)&\equiv&\langle \delta n_1\delta n_2\rangle=\nonumber\\
&=&\sum_l {(2l+1)\over 4\pi}B^2_{N,l}C^{NN}_l P_l(\cos\alpha).
\end{eqnarray}
Finally, the cross-correlation between CMB and source counts is defined as:
\begin{eqnarray}
C^{TN}(\alpha)&\equiv&\langle \delta T_1\delta n_2\rangle=\nonumber\\
&=&\sum_l {(2l+1)\over 4\pi}B_{T,l}B_{N,l} C^{TN}_l P_l(\cos\alpha)
\end{eqnarray}
with the usual definition
\begin{equation}
C_l^{TN}\equiv\langle a_{lm}^T\overline{a}_{lm}^N\rangle
\end{equation}

The theoretical auto and cross-correlation functions in a given cosmological model can be calculated by numerically integrating the Boltzmann equation for photon brightness  coupled to the other relevant equations, including the linear evolution of matter density perturbations and the evolution of gravitational potential fluctuations. We did this by suitably modifing the CMBFast code \cite{Seljak & Zaldarriaga 1996} in order to output the needed angular power spectra. In particular, we can write the angular cross-spectrum in terms of CMBFast temperature and matter transfer functions ($T_l$ and $N_l$) as:
\begin{equation}
C^{TN}_l=4\pi\int {dk\over k}\Delta^2(k)T_l(k)N_l(k)
\end{equation}
where $\Delta^2(k)\equiv k^3P(k)/2\pi^2$ and $P(k)$ is the primordial power spectrum of fluctuations. Our CMBFast modification also includes a full treatment of a generalized model of dark energy, to be described elsewhere \cite{Pietrobon & Balbi}.

\subsection{Spherical needlets}

In this paper we apply for the first time to cosmological data a new construction of spherical wavelets, the so-called {\em needlets}. Needlets were introduced into the functional analysis literature by \cite{Narcowich et al. 2006}; the investigation of their properties from the probabilistic point of view is due to \cite{Baldi et al. 2006a, Baldi et al. 2006b}, where their relevance for the statistical analysis of random fields on the sphere is pointed out for the first time. In short, the construction can be described as follows. Let $\mathcal{X}_{j}=\left\{ \xi _{jk}\right\} _{k=1,2,...,}$ denote a pixelization of the unit sphere $\mathbb{S}^{2}$; in our case we take the pixelization provided by HEALPix \cite{Gorski et al. 2005}. This pixelization divides the spherical surface in pixels of the same area, and it is hierarchical, in the sense that one can build pixelizations at higher and higher resolutions, starting from a base division of the sphere. The lowest possible resolution has 12 pixels; each subdivision increases the number of pixels in the map by a factor of 4. The level in the pixelization hierarchy is denoted by the symbol $N_{\rm side}$, with  $N_{\rm side}=1,2,4,8,...$. For each level in the pixelization hierarchy, one has 12 $N_{\rm side}^2$ pixels.  The pixelization adopted sets a limit to the highest multipole number that can be reliably extracted from a map. In our case this is given by $l=2N_{\rm side}$. We then define  $l_{\rm max}\equiv[s^{j+1}]\leq 2N_{\rm side}$ (with $[\cdot]$ denoting the integer part and $s>1$) and choose an appropriate value of $N_{\rm side}$ for the $j$ value we want to probe. 

\begin{figure}
\centering
\includegraphics[width=\columnwidth]{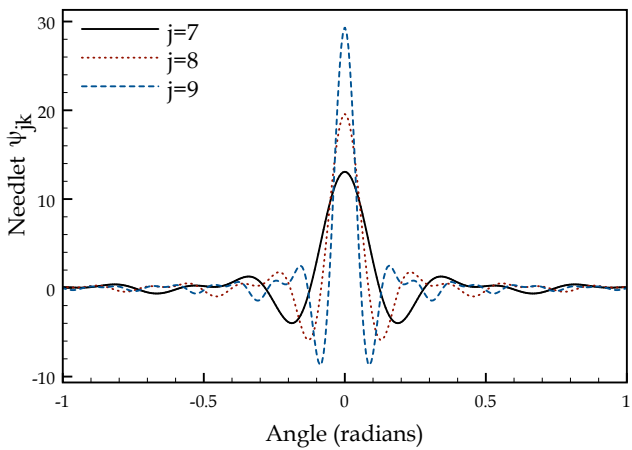}
\caption{Plots of the needlet basis $\psi_{jk}$ for an arbitrary value of $k$ and three representative values of $j$: continuous line $j=7$, dotted line $j=8$, dashed line $j=9$. Note that the angle in the $x$ axis is measured on the geodesic, not on the tangent plane. \label{fig:wave}}
\end{figure}

Let then $\Psi(\cdot)$ denote an infinitely differentiable function such that
\begin{eqnarray}
& \Psi(u) =0&\text{ for }u\leq \frac{1}{s}\text{ or }u\geq s\text{ ,} \nonumber \\
& 0 <\Psi(u)\leq 1&\text{ for }\frac{1}{s}<u<s\text{ ;}
\end{eqnarray}
The numerical implementation of such a function is described elsewhere \cite{Baldi et al. 2006b}. Up to a constant factor, we can then construct our wavelet basis as:
\begin{eqnarray}
\psi _{jk}(x)&=&\sum_{l=s^{j-1}}^{s^{j+1}}\Psi\left(\frac{l}{s^{j}}\right)P_{l}(\cos\theta) \nonumber \\
&=&\sum_{l=s^{j-1}}^{s^{j+1}}\Psi\left(\frac{l}{s^{j}}\right)\sum_{m=-l}^{l}\overline{Y}_{lm}(x) Y_{lm}(\xi _{jk}),
\end{eqnarray}
where $\theta$ denotes the angle between the two vectors $x,\xi _{jk}$. Some representative cases for the wavelet basis $\psi _{jk}$ are plotted in Figure~\ref{fig:wave}. The wavelets coefficients are then derived as 
\begin{eqnarray}
\beta _{jk} &=&\int_{\mathbb{S}^{2}}T(x)\psi _{jk}(x)dx \nonumber \\
&=&\int_{\mathbb{S}^{2}}\left\{ \sum_{l=1}^{\infty
}\sum_{m=-l}^{l}a_{lm}Y_{lm}(x)\right\} \psi _{jk}(x)dx \nonumber \\
&=&\int_{\mathbb{S}^{2}}\sum_{l=1}^{\infty
}\sum_{m=-l}^{l}a_{lm}Y_{lm}(x)\times \nonumber\\
&\times&\left\{ \sum_{l^{\prime }=\left[ s^{j-1}
\right] }^{\left[ s^{j+1}\right] }\Psi\left(\frac{l^{\prime }}{s^{j}}\right)\sum_{m^{\prime }=-l^{\prime }}^{l^{\prime }}\overline{Y}_{l^{\prime }m^{\prime }}(x)Y
_{l^{\prime }m^{\prime }}(\xi _{jk})\right\} dx \nonumber\\
&=&\sum_{l^{\prime }=\left[ s^{j-1}\right] }^{\left[ s^{j+1}\right] }\Psi\left(\frac{
l^{\prime }}{s^{j}}\right)\sum_{l=1}^{\infty }\sum_{m=-l}^{l}a_{lm}\times\nonumber\\
&\times&\sum_{m^{\prime
}=-l^{\prime }}^{l^{\prime }}Y_{l^{\prime }m^{\prime }}(\xi
_{jk})\left\{ \int_{\mathbb{S}^{2}}Y_{lm}(x)\overline{Y}_{l^{\prime }m^{\prime }}(x)dx\right\}  \nonumber\\
&=&\sum_{l=\left[ s^{j-1}\right] }^{\left[ s^{j+1}\right] }\Psi\left(\frac{l}{s^{j}}
\right)\sum_{m=-l}^{l}a_{lm}Y_{lm}(\xi _{jk})\text{ ;}
\end{eqnarray}
here again we have used $[\cdot]$ to denote the integer part of a real number. 

In our view, the needlets coefficients $\beta _{jk}$ enjoy a number of very important properties which are not shared by other spherical wavelets. As a first example, we stress their localization in the multipole space; the contribution from the $\left\{ a_{lm}\right\} $ is analytically derived to be bounded between the multipoles $l=\left[ s^{j-1}\right] $ and $l=\left[s^{j+1}\right] .$ In contrast with some of the existing literature, here the dependence is sharp and explicit on the user-chosen parameter $s.$ A second important property concerns localization properties in pixel space; indeed, it is shown in \cite{Narcowich et al. 2006} that
\begin{eqnarray}
|\psi _{jk}(x)|&\leq& \frac{C_{M}s^{j}}{(1+s^{j}d(x,\xi _{jk}))^{M}},\nonumber\\
&&\text{for all }x\in \mathbb{S}^{2},\text{ }M=1,2,3,\dots
\end{eqnarray}
where $d(x,\xi _{jk})$ denotes angular distance, $M$ is any positive integer and $C_{M}$ is a constant which does not depend either on $x$ or on $j.$ To be more explicit, this result implies that, outside any neighbourhood of radius $\varepsilon $ around the direction $\xi _{jk}$ ($N_{\varepsilon
}(\xi _{jk}),$ say), we have 
\begin{eqnarray}
|\psi _{jk}(x)|&\leq& \frac{C_{M}}{(s^{j}\varepsilon )^{M-1}},\nonumber\\
&&\text{for all 
}x\notin N_{\varepsilon }(\xi _{jk}),\text{ }M=1,2,3,\dots
\label{narco}
\end{eqnarray}
that is, the decay of the wavelet transform is faster than any polynomial $s^{j}.$ This clearly establish an excellent localization behaviour in pixel space. Note that the constants $C_{M}$ do depend on the form of the weight function $\Psi$, and in particular on the value of the bandwidth parameter $s$; typically a better localization in multipole space (i.e., a value of $s$ very close to unity) will entail a larger value of $C_{M}$, that is, less concentration in pixel space for any fixed $j$.

The probabilistic properties of the random coefficients $\beta _{jk}$ have been established in \cite{Baldi et al. 2006b}; in that paper, it is shown that for any two (sequence of) pixels $\xi _{jk},\xi _{jk^{\prime }}$ such that their angular distance is larger than a positive $\varepsilon ,$ for all $j$, we have
\begin{equation}
\frac{\langle\beta _{jk}\beta _{jk^{\prime }}\rangle}{\sqrt{\langle\beta
_{jk}^{2}\rangle\langle\beta _{jk^{\prime }}^{2}}\rangle}\leq \frac{C_{M}}{(s^{j}\varepsilon )^{M-1}}\text{ for all }M=1,2,3,\dots
\label{baldi}
\end{equation}
thus proving wavelets coefficients are asymptotically uncorrelated as $j\rightarrow \infty $ for any fixed angular distance; equation (\ref{baldi}) is clearly a probabilistic counterpart of (\ref{narco}). To the best of our knowledge, this is the first example of such kind of results for any type of spherical wavelets: asymptotic uncorrelation (i.e.\ independence in the Gaussian case) simplifies enormously any statistical inference procedure. In particular, equation (\ref{baldi}) is used in \cite{Baldi et al. 2006b} to derive analytically the asymptotic behaviour of a number of procedures based on needlets, including tests on angular power spectra or tests for Gaussianity and isotropy. In a similar manner, it is possible to show that needlets coefficients outside the masked regions are asymptotically unaffected by the presence of incomplete sky coverage. 

\begin{figure}
\centering
\includegraphics[angle=90,width=0.8\columnwidth]{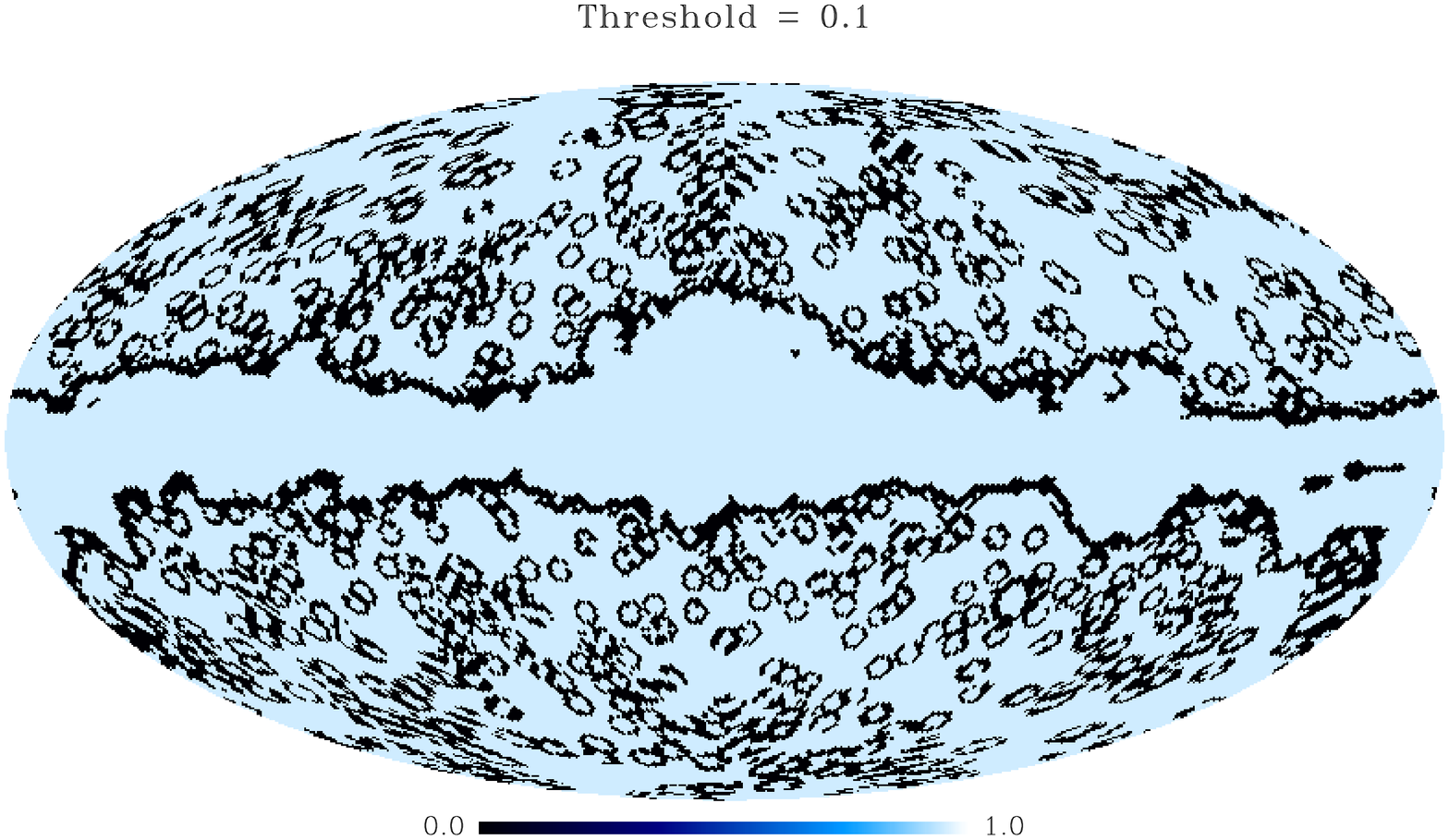}
\includegraphics[angle=90,width=0.8\columnwidth]{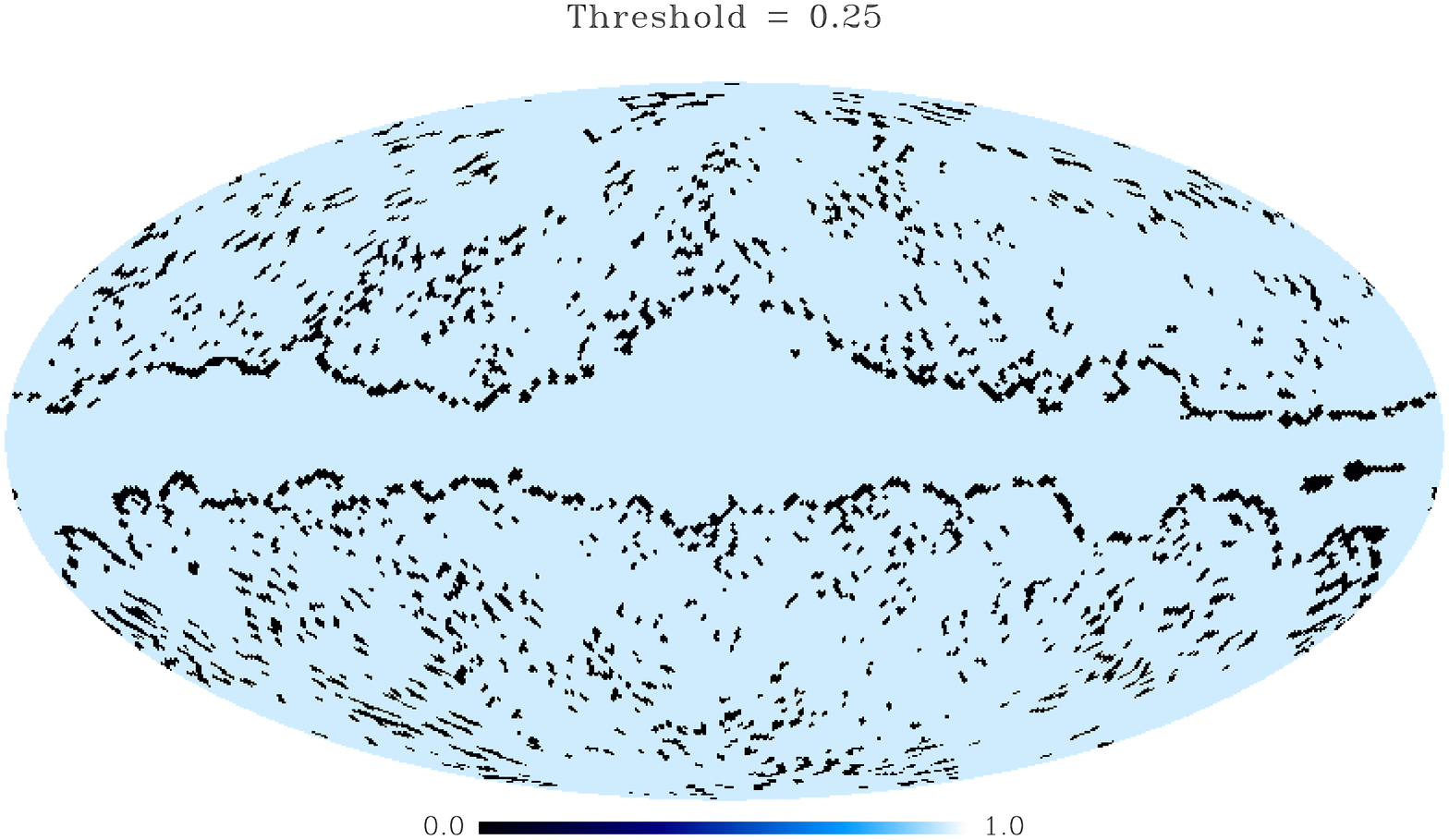}
\includegraphics[angle=90,width=0.8\columnwidth]{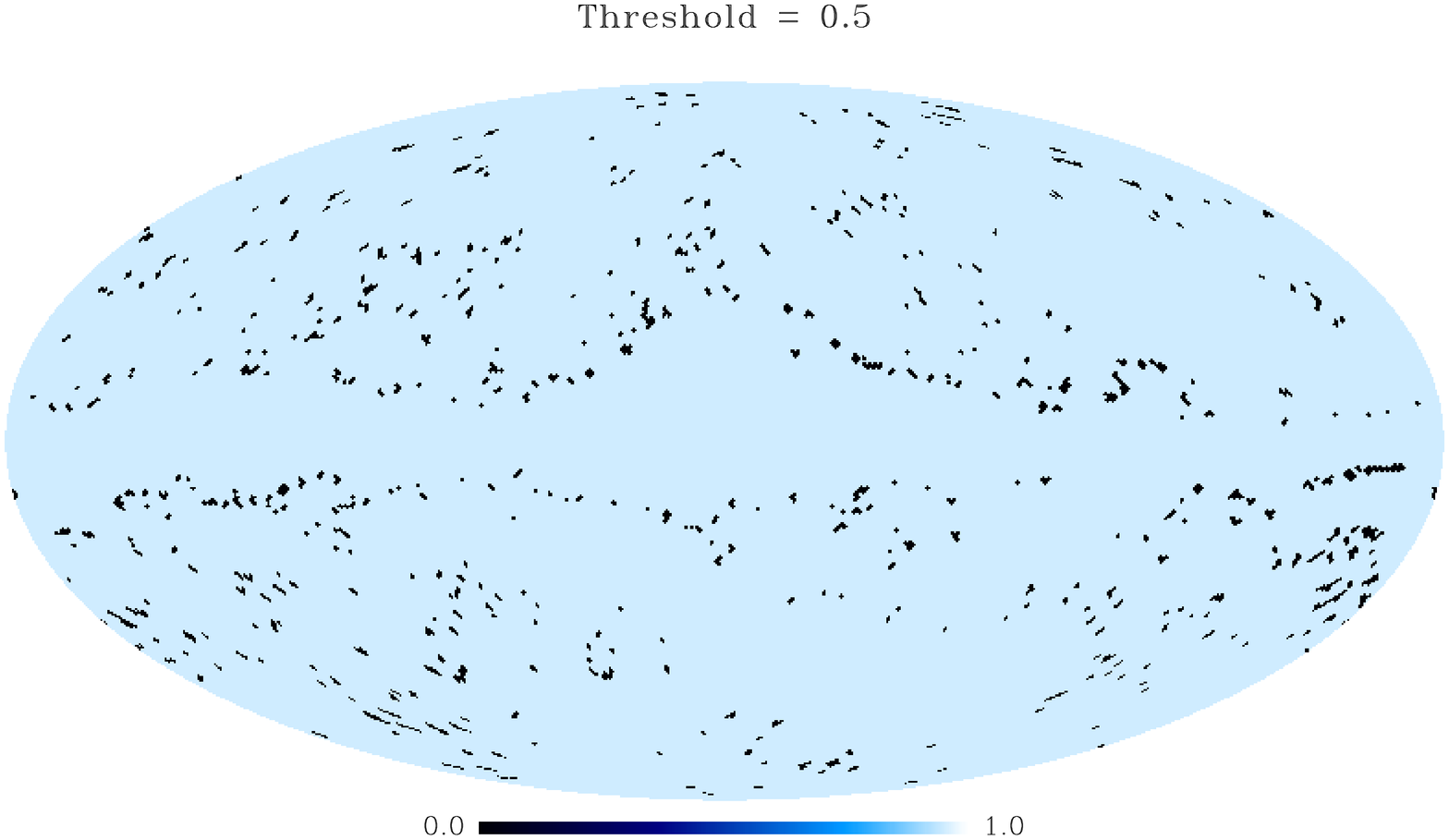}
\caption{Effect of the sky cut on the estimated needlet coefficients. Pixels $k$ corresponding to values of the estimator $D_{jk}$ above a given threshold are shown in black, and are clearly concentrated only very close to the masked regions. For this figure, $j=11$. From top to bottom, the threshold for $D_{jk}$ takes the values 0.1, 0.25, 0.5. \label{fig:needlet}}
\end{figure}

To provide some numerical evidence of this localization property of the needlets coefficients we evaluate, for each pixel outside the Kp0 mask, a normalized squared error by taking 
\begin{equation}
D_{jk}= \langle(\beta_{jk}-\hat{\beta}_{jk})^{2}\rangle/\langle\beta_{jk}^2\rangle, \label{nd}
\end{equation}
where $\hat{\beta}_{jk}$ is the wavelet coefficient calculated in presence of a sky mask. More explicitly, $D_{jk}$ represents a measure of the difference between the needlets coefficients evaluated with and without partial sky coverage. In the presence of good localization properties in pixel space, we should expect this difference to be small outside some neighbourhoods of the masked regions. This expectation is indeed confirmed by our simulation results, which are based upon 100 replicated temperature maps. More precisely, we observe that in 64\% of pixel outside the masked region $D_{jk}$ took values smaller than 0.1; this percentage rises to 88\% when the threshold value is increased to 0.25 and to 97\% when we take 0.5 as a threshold value. To add some visual evidence as a support to our claims, pixels above the indicated threshold are plotted in Figure~\ref{fig:needlet}; their concentration at the boundary of the masked region (i.e.\ next to the galactic plane, and around point sources) confirms the rationale beyond our claims. These issues shall be further investigated in an upcoming paper; see also \cite{Baldi et al. 2006a, Baldi et al. 2006b}.

As a final advantage, we stress that our construction is explicitly devised for the sphere, and as such it does not rely on any sort of tangent plane approximation; we view this as an important asset both from the computational point of view and in terms of the accuracy of statistical results.

Having extracted the needlets coefficients $\beta_{jk}$ from the CMB and source count maps, the cross-correlation estimator in wavelet space, $\beta_j$, can be calculated simply as:
\begin{equation}\label{betaj}
\beta_j\equiv\sum_k {1\over N_{\rm pix}(j)}\beta_{jk}^T\overline{\beta}_{jk}^N
\end{equation}
where $N_{\rm pix}(j)$ is the number of pixels in the pixelization scheme (given by $N_{\rm pix}=12 N_{\rm side}^2$).
The theoretical prediction for $\beta_j$ can be computed from the expected $C_l^{TN}$ as:
\begin{equation}
\beta_j=\sum_l {(2l+1)\over 4\pi}\,\left[\Psi\left(\frac{l}{s^{j}}\right)\right]^2B_{T,l}B_{N,l}C_l^{TN}
\end{equation}

It can be easily shown that the analytical expression for the dispersion of the estimated cross-correlation power spectrum in needlet space is:
\begin{equation}\label{errors}
\Delta \beta_j =\left( \sum_l \frac{(2l+1)}{16\pi^{2}}  \left[\Psi\left(\frac{l}{s^{j}}\right)\right]^4 \left(\left(C^{TN}_l\right)^{2} + C^T_l C^N_l\right)\right)^{1/2}
\end{equation}
which, of course, must be only taken as an approximation when dealing with real data.

\section{Results}
\label{results}

\begin{figure}
\centering
\includegraphics[width=\columnwidth]{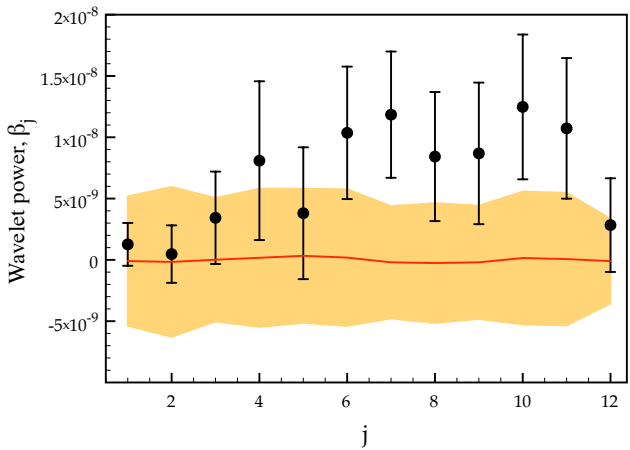}
\caption{The wavelet cross-correlation power spectrum $\beta_j$ of the WMAP and NVSS maps. The points represent the signal extracted from the real data, with error bars given by Eq. \ref{errors}. The continuos line is the average of the cross-correlation power spectra obtained when 1000 simulated CMB fiducial data sets are correlated with the real NVSS map: this measures the level of correlation expected from casual alignment. The shaded area is the $1\sigma$ dispersion of the simulated spectra. \label{fig:signal}}
\end{figure}

\subsection{ISW detection}

In Figure \ref{fig:signal} we show the cross-correlation signal in wavelet space extracted from the WMAP and NVSS data. The data points shown in the Figure were obtained following the above described procedure, applying Equation \ref{betaj}. We chose the value $s=1.5$ in the wavelet construction for our analysis. The excess signal peaks at value $7<j<11$, corresponding to angular scales between $2^\circ$ and  $10^\circ$, as expected from theoretical studies \cite{Afshordi 2004}. The error bars were calculated according to Equation \ref{errors}. In order to check that the observed signal was not produced by casual alignment of sources in the NVSS catalogue with the CMB pattern at decoupling, we produced $1000$ Monte Carlo simulations of the CMB sky with an underlying theoretical fiducial lambda cold dark matter (LCDM) model corresponding to the WMAP 3 year best fit. The resulting maps were processed through our analysis pipeline, and the cross-correlation with the real NVSS map was calculated for each simulated data set. Figure \ref{fig:signal} shows the resulting average cross-correlation signal (continuous line), which is basically zero on all scales. The standard deviation of the simulations is also shown in the same Figure (shaded area). These errors, calculated through the Monte Carlo procedure, are consistent with the analytical estimates of Equation \ref{errors}. 

The cross-correlation signal extracted from the data is significantly higher than the expectation value of the simulated data. To quantify the statistical significance of the detection, we computed the quantity $X^2\equiv\sum_j (\beta^{\rm obs}_j-\langle\beta^{\rm sim}_j\rangle)^2/\Delta\beta_j^2$. We found $X^2=29.8$. Assuming $X^2$ is distributed as a $\chi^2$ with $12$ degrees of freedom (corresponding to our 12 data points), we can exclude that the $\beta_j$ were produced under the null hypothesis with $99.7$\% confidence. This result is in agreement with previous studies which used the same data \cite{Nolta et al. 2004, Vielva et al. 2006, McEwen et al. 2006}. We stress again the fact that in our analysis the very nature of needlets guarantees that the correlation between adjacent data points is very low, even in the presence of sky cuts.

\subsection{Consequences for dark energy models}

\begin{figure}
\centering
\includegraphics[angle=90,width=\columnwidth]{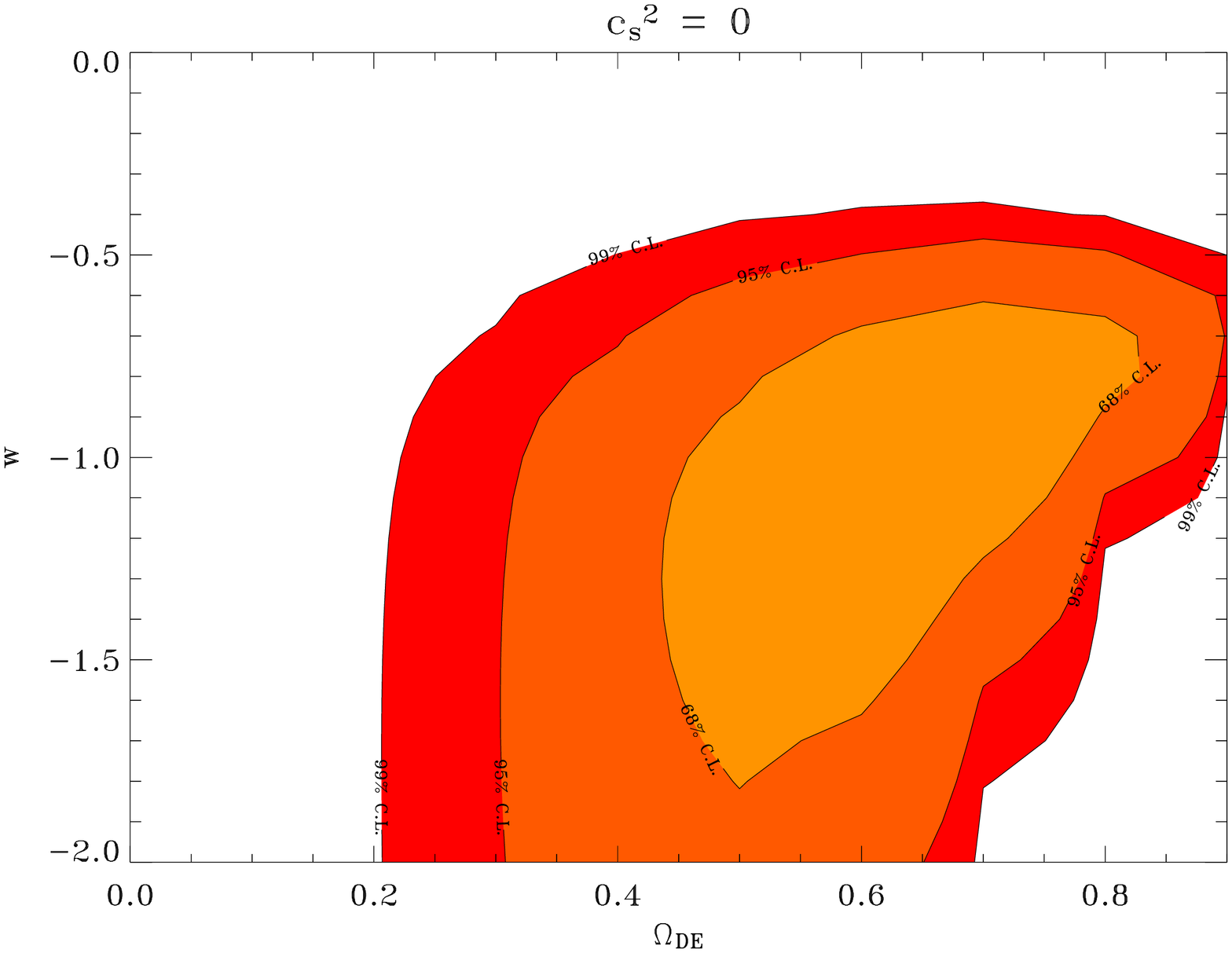}
\includegraphics[angle=90,width=\columnwidth]{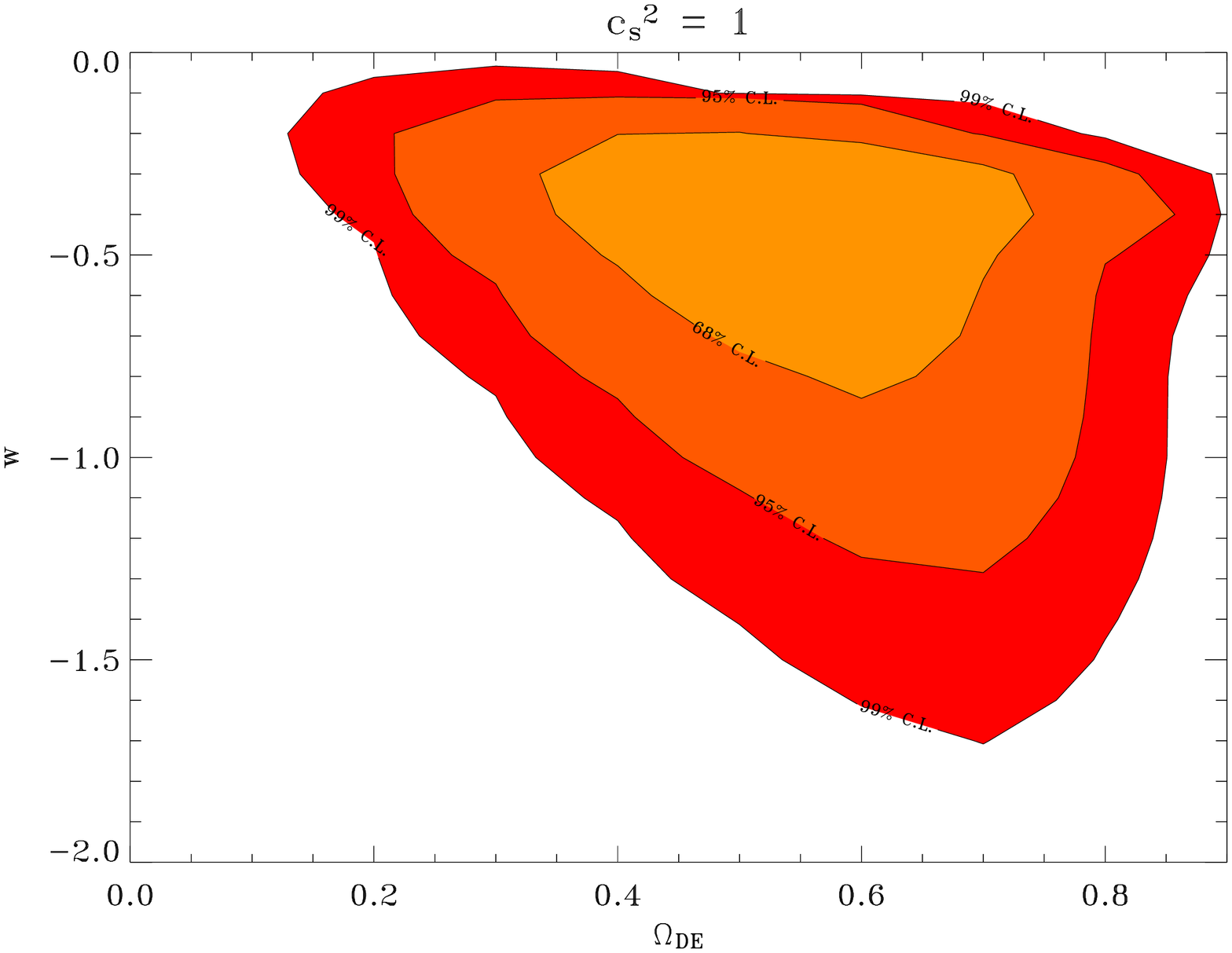}
\caption{Constraints at 68\%, 95\% and 99\% confidence level in the $\omde$---$w$ plane. The upper panel was obtained under the hypothesis that the dark energy speed of sound is $\cs2=0$; the lower panel was obtained for $\cs2=1$. \label{fig:contours}}
\end{figure}

\begin{figure}
\centering
\includegraphics[width=\columnwidth]{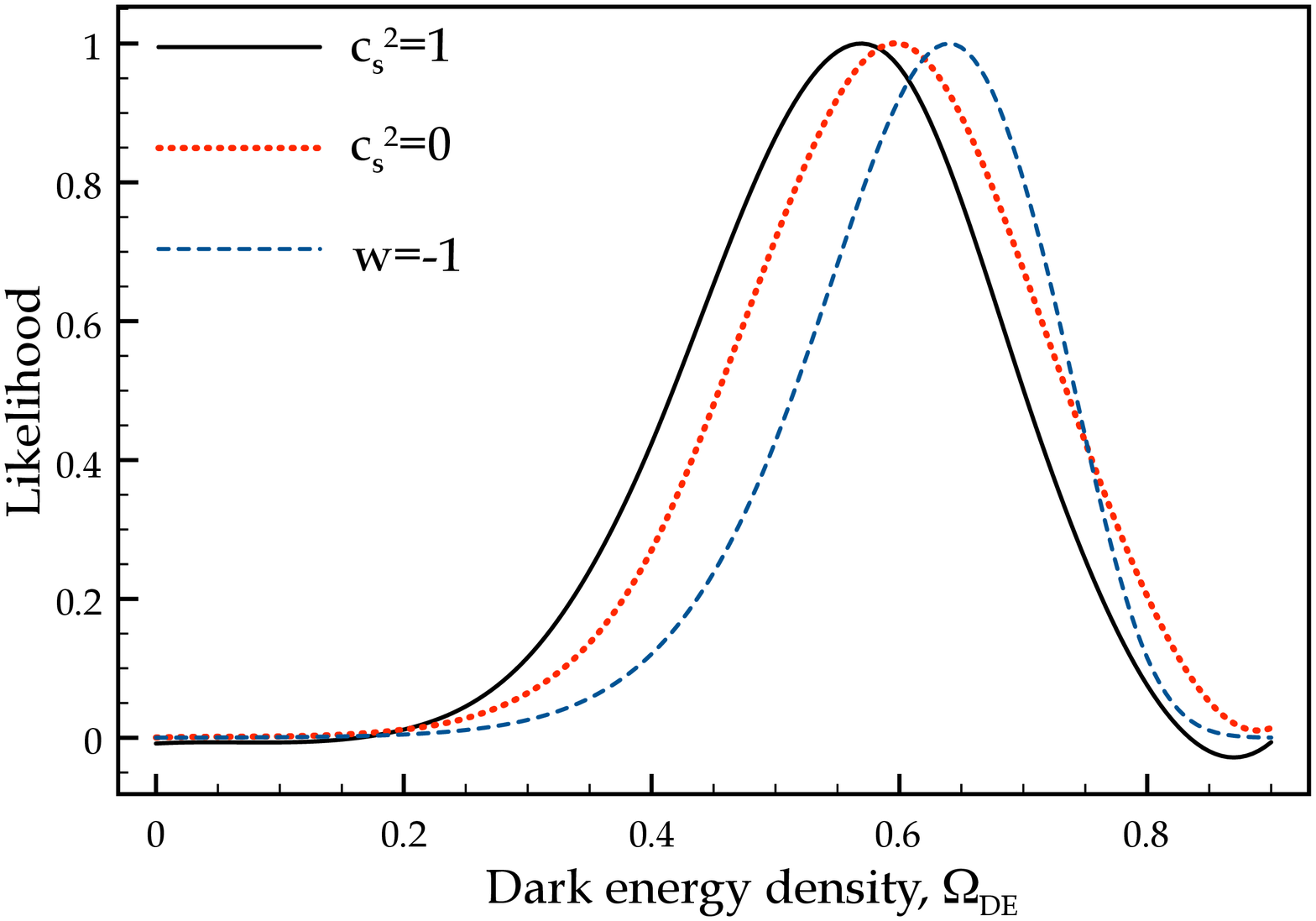}
\includegraphics[width=\columnwidth]{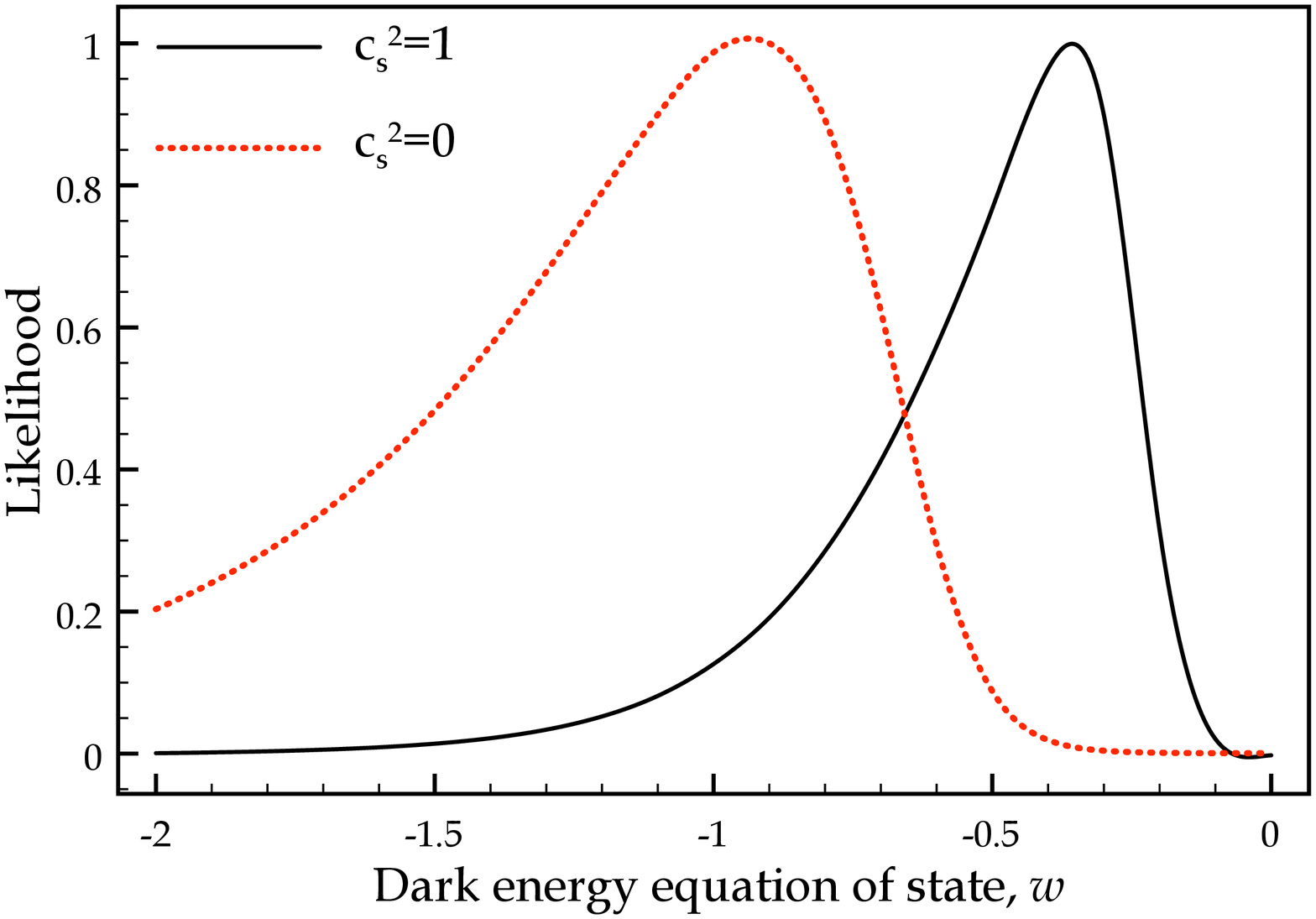}
\caption{Marginalized likelihood function for the dark energy density $\omde$ (upper panel) and equation of state $w$ (lower panel). In each panel, the continuous curve was obtained under the hypothesis that $\cs2=1$, while the dotted curve is for $\cs2=0$. \label{fig:like}}
\end{figure}

\begin{figure}
\centering
\includegraphics[width=\columnwidth]{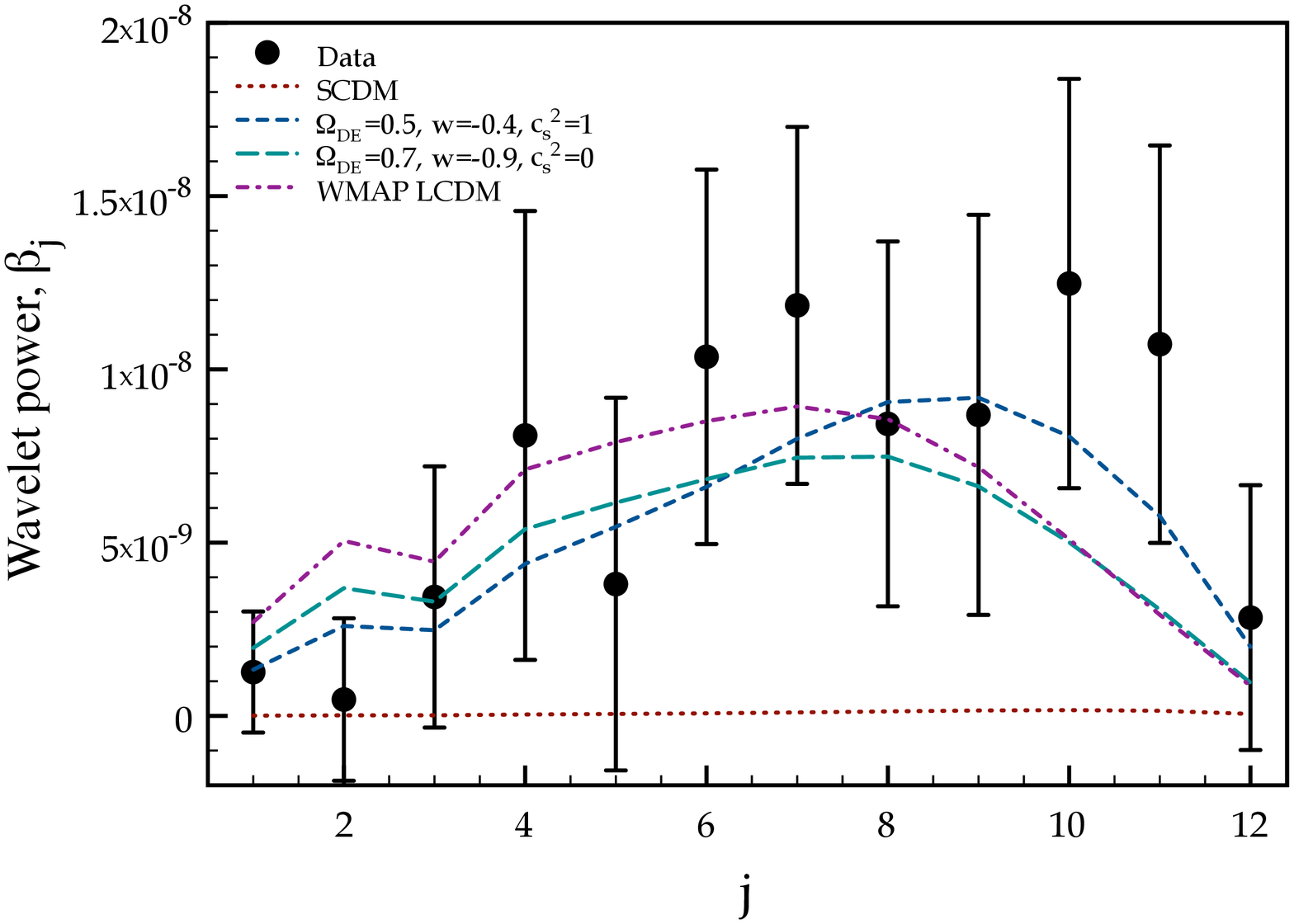}
\caption{The cross-correlation data points estimated in our analysis (big dots), with their errors, are shown together with the theoretical predictions for a standard CDM model (without dark energy, dotted curve), the best fit under the hypothesis that $\cs2=1$ (short-dashed curve), the best fit under the hypothesis that $\cs2=0$ (long-dashed curve) and the LCDM ($w=-1$) that best fits the WMAP 3 year data (dot-dashed curve). \label{fig:models}}
\end{figure}

We have compared the cross-correlation estimated from the data with the theoretical predictions in order to set constraints on dark energy models. Following the phenomenology described in \cite{Hu 1998}, we have modeled the dark energy component as a fluid characterized by its density parameter, $\omde$,  its equation of state, $w=p/\rho$, and the sound speed $c_s^2= \delta p/\delta \rho$. The latter, needs not be the usual adiabatic one, but also accounts for an entropic sound speed, so that $c_s^2\neq\dot p/\dot\rho$.
We have considered the above three quantities as the free parameters of our analysis. All the other cosmological parameters were held fixed at the best fit values estimated using the WMAP 3 year data: $\Omega=1$, $H_0=73$ km/s/Mpc, $\Omega_{\rm b}=0.042$, $\tau=0.088$, $A=0.68$ (or $\sigma_8=0.74$), $n_s=0.951$ \cite{Spergel et al. 2006}. When modeling the NVSS catalogue, we adopted a constant bias parameter $b=1.6$, a value suggested in \cite{Dunlop & Peacock 1990} and already used in previous analysis, e.g.\ \cite{Boughn & Crittenden 2004, Boughn & Crittenden 2005, Nolta et al. 2004, Vielva et al. 2006}. 

We have computed the theoretical expectation for the cross-correlation by using Equation~\ref{betaj} and the output of our modified CMBFast code \cite{Pietrobon & Balbi}. Our calculation fully takes into account the clustering properties of dark energy. We restricted our analysis to two different values of $\cs2$ corresponding to the limiting cases describing a scalar field behaviour ($\cs2=1$) and a matter behaviour ($\cs2=0$). 

The main results of our analysis are summarized in Figure~\ref{fig:contours} where we plot the joint constraints on the dark energy parameters $\omde$ and $w$ for the two cases of sound speed considered here, and in Figure~\ref{fig:like}, where we show the separate marginalized likelihoods for $\omde$ and $w$.

The first conclusion we can draw from our analysis is that the evidence for non zero dark energy density is rather robust: we find $0.32\leq\omde\leq 0.78$ for $\cs2=1$ and $0.36\leq\omde\leq 0.81$ for $\cs2=0$, both at 95\% confidence level. A null value of $\omde$ is excluded at more than $4\sigma$ (see Figure~\ref{fig:like}, upper panel), independently of $\cs2$. When we model the dark energy as a cosmological constant (i.e.\ we assume the value $w=-1$ for its equation of state), the bounds on its density shrinks to $0.41\leq\omde\leq 0.79$ at 95\% confidence level. 

On the other hand, the constraints on $w$ are strongly influenced by the assumed value of $\cs2$, because of the different clustering behaviour of dark energy (Figure~\ref{fig:like}, lower panel). In particular, we find that the value of the equation of state which corresponds to a cosmological constant ($w=-1$) is well within the 95\% c.l.\ when we assume $\cs2=0$. In this case, we can only put an upper bound at 95\% c.l.: $w\leq -0.54$. When $\cs2=1$ is assumed, we find that phantom models are excluded and that the cosmological constant case performs comparatively worse than models with larger values of $w$. Our bounds at 95\% c.l.\ are $-0.96\leq w \leq -0.16$. However, we emphasize that, for values of $\omde\sim 0.7$, models with $w=-1$ are a good fit to the data, as it is evident from Figure~\ref{fig:contours} (lower panel). In fact, the LCDM WMAP best fit (with $\omde=0.76$ and $w=-1$) has $\chi^2=9.35$, with 12 data points. 
The predicted cross-correlation for some dark energy models is shown together with our data points in Figure~\ref{fig:models}.

\section{Conclusions}
\label{conclusions}

We have analyzed the WMAP 3 year CMB temperature data, in conjunction with the NVSS radio galaxy survey, and found further evidence of a correlation between the CMB fluctuation pattern and the local distribution of matter, consistent with an ISW effect taking place at a late epoch of cosmic evolution. When a flat universe is assumed (as suggested by CMB observation) the detection of a late ISW signature is a strong evidence in favour of a dark energy component.  Our findings are based on a new construction of spherical wavelets that has a number of advantages with respect to previous studies. The presence of a correlation between the CMB and the LSS is established with a high level of confidence. 

We have also improved the treatment of the dark energy component, introducing a more general parameterization than those used is similar earlier analysis. Quite interestingly, we find that although the case for a non zero dark energy contribution to the total density is compelling and robust, the constraints on $w$ do depend on the assumed clustering properties of the dark energy component, namely its sound speed $\cs2$. Phantom models, and also the ordinary cosmological constant case $w=-1$, perform worse when a quintessence behaviour $\cs2=1$ is assumed. This is due to the fact that there exist models with $w\sim -0.4$ which predict more correlation at smaller angular scales ($\theta\sim 2^\circ$). This is an intriguing result, that could imply a ISW effect taking place at redshifts as high as $z\sim 1$, earlier than expected in the cosmological constant case.  A similar preference for larger values of $w$ in quintessence models was also found in \cite{McEwen et al. 2006}.

Whether this is an indication of interesting physics taking place between the dark energy and dark matter components is a subject that requires further investigation. Clearly, the observation of ISW is proving quite promising as a tool to answer the questions arising from the mysterious nature of dark energy. While the CMB data have reached a great degree of accuracy on the angular scales that are more relevant for the detection of ISW, deeper redshift surveys and better catalogues can, in the future, improve the tracing of the local matter distribution, thus allowing to reduce the errors on the cross-correlation determination.

\acknowledgements
We wish to thank Paolo Natoli for several useful discussions. D.~M.  is especially grateful to P. Baldi, G. Kerkyacharian and D. Picard. D.~P.\ is supported by INAF.

\end{document}